\documentclass[aps,floatfix,prstab,preprint,tightenlines,superscriptaddress,showpacs]{revtex4}
\usepackage{amsfonts}
\usepackage{graphicx}
\usepackage{color}
\begin{document}


\title{Black-body radiation in Tsallis statistics}
\author{H. Chamati}
\email{chamati@issp.bas.bg}
\affiliation{Institute of Solid State Physics,
72 Tzarigradsko Chauss\'ee, 1784 Sofia, Bulgaria}
\author{A.Ts. Djankova}
\affiliation{Institute of Applied Physics,
Technical University, 8 Kliment Ohridski St.,
1000 Sofia, Bulgaria.}
\author{N.S. Tonchev}
\email{tonchev@issp.bas.bg}
\affiliation{Institute of Solid State Physics,
72 Tzarigradsko Chauss\'ee, 1784 Sofia, Bulgaria}

\begin{abstract}
Some results for the black-body radiation obtained in the context
of the $q$-thermostatistics are analyzed on both thermodynamical
and statistical-mechanical levels. Since the thermodynamic
potentials can be expressed in terms of the Wright's special
function an useful asymptotic expansion can be obtained. This
allows the consideration of the problem away from the
Boltzmann-Gibbs limit $q=1$. The role of non-extensivity, $q<1$,
on the possible deviation from the Stefan-Boltzmann $T^{4}$
behavior is considered. The application of some approximation
schemes widely used in the literature to analyze the cosmic
radiation is discussed.
\end{abstract}
\maketitle

\section{Introduction}\label{introduction}
The nonextensive statistical mechanics (NSM) is based on the
generalized entropy defined by:
\begin{equation}\label{entropy}
S^{T}_q=-k\frac{1-\sum_{q=1}^W p_i^q}{1-q},
\end{equation}
where the index $i$ labels the possible microstates of the system
under consideration, $\{p_i\}$ is a set of normalized
probabilities, the real parameter $q$ characterizes the degree of
nonextensivity and $k$ is a positive constant. Notice that taking
the limit $q\to1$ leads to the popular Boltzmann-Gibbs statistics
(BGS). For a recent review on the nonextensive thermostatistics
and its current status see Refs. \cite{TB2004,tsallis2004}.

The thermodynamics in the context of the (NSM) is
investigated by generalizing the Gibbs canonical ensemble to the
case $q\neq1$. This is achieved by maximizing the entropy defined
by Eq. (\ref{entropy}) under the constrains: (i) normalization of the
probabilities, and (ii) knowledge of the expectation value of the energy. 
The expectation values that lie in basis
of the thermostatistical considerations are usually computed using two
different approaches. The first one is the so the called
`unnormalized' approach proposed in Ref. \cite{CT91}. Within this
approach, for a given observable $O$ with an eigenvalue $O_i$ in
the microstate $i$ one has
\begin{equation}\label{CT}
\langle O\rangle = \sum_{i=1}^W p_i^q O_i.
\end{equation}
This approach shows several difficulties in describing the
thermodynamics (see e.g. \cite{tsallis2004}). It is unable to preserve
many of the thermodynamic properties. To overcome this inconveniences the
`normalized approach' has been advanced in Ref. \cite{TMP98}, where the
expectation values are given by
\begin{equation}\label{OLM}
\langle O\rangle = \frac{\sum_{i=1}^W p_i^q O_i}
{\sum_{i=1}^W p_i^q}.
\end{equation}
The normalized treatment seems to provide one with a natural
bridge that connects the NSM to the thermodynamics
\cite{tsallis2004}. The normalized approach has been in turn
improved by so called `optimal Lagrange multipliers' (OLM)
approach \cite{MNPP02}. Nowadays it is believed that both
approaches are the most appropriate choice for investigating the
thermodynamics within the framework of the NSM regarding the
nature of the Lagrange multiplier associated with the temperature.
In these cases the nonextensivity is restricted just to the
entropy, while the internal energy remains extensive as in the
case of the BGS. The Lagrange multipliers preserve their
traditional intensive character and can be identified with their
thermodynamic counterparts. An important consequences of the
success of both approaches is the unification of the Tsallis and
R\'enyi variational formalism under a common context. This success
is originating from the fact that the R\'enyi's entropy is
extensive\cite{tsallis2004}. Let us mention that the extensivity
of the internal energy is true as long as we consider a system
with a large number of particles or the thermodynamic limit under
the condition $q<1$ \cite{abe1999}.

In the last few years many papers have been published on the
application of NSM to the black-body radiation
\cite{TBL95,PPV,LM98,WNM,TT00,AT01,MPPTa,MPPT,BSD02}. Our belief
is that rigorous and exact results have an instructive role in the
field. The exact expression for the corresponding partition
function has been obtained in Refs. \cite{LM98,MPPTa,MPPT}. The
studies presented in Refs.
\cite{TBL95,PPV,LM98,WNM,TT00,AT01,BSD02} employ the unnormalized
approach, while those in Ref. \cite{MPPTa,MPPT} used the
normalized one.

Irrespective of the used approach one can see that in the case
$q\neq1$
the derivation of exact results is a
very complicated task. The final expressions one has at hand
are too cumbersome that obscures the underlying physics. In this
situation an approximation  that tries to make the exact results
more simple and transparent is preferable. However, the real
benefit from the exact treatment without well defined range of
validity of the used approximation seems to be doubtful. Let us
note that the more appropriate approximations are limited to
simply computing $(1-q)$ corrections (see Refs. \cite{TT00} and
\cite{MPPT} and references therein) since the Boltzmann-Gibbs
limit $q\rightarrow 1$ leads to great simplifications. In this
situation the possible strong deviations from the usual
Boltzmann-Gibbs case are of a significant interest. Quite recently
the necessity to have a well estimated approximation in this field
grows up, since some strong
 criticism concerning the physical validity of Tsallis statistics
takes place \cite{N03,Ncm}. The objections of Refs. \cite{N03,Ncm}
in their major parts are concentrated on applications of
$q$-thermostatistics to the black-body radiation. Recently
this issue
has been a matter of a debate in the literature \cite{N03,Ncm,T03}.

The aim of the present study is to illustrate another
possibility for simplification of the basic expressions in both
approaches not related to the small value of $(1-q)$. It is
based on the fact that in both cases the intricate sums that
appear in the theory may be presented \cite{ASL} in terms of the
Wright function with well studied analytical properties
\cite{GLM99}. This is justified since we are considering a tremendous
system. We hope this possibility will shed some light on the
existing debate \cite{N03,Ncm,T03}.

This paper is organized as follows: In Section \ref{sec2} we discuss
the thermodynamic derivation of the popular Stefan-Boltzmann law in
terms of the Tsallis statistics. In Section \ref{wrightsec} we introduce
the mathematical background we need in our analysis. This is presented in
Section \ref{secsb}. Section \ref{discussion} is devoted to the
discussion of our results.

\section{Some thermodynamic relations}\label{sec2}
First we shall introduce some basic notions. The radiation field
in a large cavity can be considered to consist of a denumerably
infinite set of electromagnetic oscillators corresponding to the
various quantum states ${\mathbf k}$ in a $d$-dimensional box. The
oscillator frequencies, $\omega_{i}=ck_{i}$, are related to the
total energy $E$ by $E=\Sigma_{i}n_{i,\epsilon}\hbar\omega_{i}$,
where $n_{i,\epsilon}$ is the number of oscillator quanta with
frequency $\omega_{i}$ and polarization $\epsilon$, $c$ is the
light  speed, $\hbar$ is the Plank constant and $k_{i}=|{\mathbf
k}_{i}|$. The Boltzmann-Gibbs partition function $Z_{1}$, for a
large volume $V$, can be written as
\begin{equation}\label{1}
Z_{1}(A_{d})=\exp(A_{d}),
\end{equation}
where
\begin{equation}\label{2}
A_{d}=\frac{\Gamma(d)\zeta(d+1)2\tau_{d}}{(4\pi
)^{d/2}\Gamma(d/2)} \left(\frac{k_BT}{\hbar c}\right)^{d}V.
\end{equation}
In (\ref{2}), $\tau_{d}=d-1$ is the number of linear-independent
polarizations, $k_B$-Boltzmann constant, $T$-temperature, and
$\Gamma(x)$ and $\zeta(x)$ are Gamma and Zeta function,
respectively.

The main obstacles related to the applicability of the
$q$-thermostatistics to the black-body radiation may be considered
in the context of the famous Stefan-Boltzmann law. In the
Boltzmann-Gibbs thermodynamics the Stefan-Boltzmann law follows
from the equation
\begin{equation}\label{slt}
\left(\frac{\partial U}{\partial
V}\right)_{T}=T\left(\frac{\partial p}{\partial T}\right)_{V}-p
\end{equation}
and the relation
\begin{equation}\label{eos}
p(T)=\frac{u(T)}d,
\end{equation}
where $p\equiv p(T)$ is the pressure and $u(T)=U(T,V)/V$ - the 
internal energy per
unit volume. Here the dependence on the temperature
alone is crucial. As a result Eq. (\ref{slt}) reduces to an
ordinary differential equation for $u(T)$ and its solution is
$u(T)=\sigma T^{d+1}$, where $\sigma$ is a constant that cannot
be obtained on the macroscopic level.

Let us now consider the corresponding generalization of the
Stefan-Boltzmann law to the NSM context. In the $q$-thermodynamics the
following expressions for the internal energy $U_{q}(T,V)$ and the pressure
$p_{q}(T,V)$ hold \cite{LM98}:
\begin{equation}\label{U}
U_{q}(T,V)=kT^{2}\frac{\partial}{\partial
T}\frac{[Z_{q}]^{1-q}-1}{(1-q)}
\end{equation}
and
\begin{equation}\label{P}
p_{q}(T,V)=kT\frac{\partial}{\partial
V}\frac{[Z_{q}]^{1-q}-1}{(1-q)},
\end{equation}
where $Z_{q}$ is the $q$-generalized partition function. Now, we shall
give some thermodynamic relations for the black-body radiation using as
input the definitions (\ref{U}) and (\ref{P}).  Because of the simple
dimensional arguments it is evident that $Z_{q}\equiv Z_{q}(A_{d})$. If we
introduce the convenient notation $\ln_q x=\frac{x^{q-1}-1}{1-q}$ the
internal energy $U_{q}(T,V)$ may be expressed through
$Z_{q}(A_{d})$ and
\begin{equation}\label{5a}
U_{q}(T,V)= dkTA_{d}\frac{\textrm{d}}{\textrm{d}A_{d}}
\ln_qZ_q(A_d).
\end{equation}
Correspondingly for the pressure $p_{q}(T,V)$ we get
\begin{equation}\label{5aa}
p_{q}(T,V)V=kTA_{d}\frac{\textrm{d}}{\textrm{d} A_{d}}
\ln_qZ_q(A_d).
\end{equation}
From Eqs. (\ref{5a}) and (\ref{5aa}) immediately follows the
$q$-generalization of the relation (\ref{eos})
\begin{equation}\label{imp}
p_{q}(T,V)V=\frac{U_{q}(T,V)}{d}.
\end{equation}
The relation (\ref{imp}) between the pressure and internal energy is
$q$-independent, as it should be. The violation of the relation
between the pressure and the internal energy would compromise the
theory since this can be established from pure electrodynamic reasoning.
Eq. (\ref{imp}) was verified in Refs. \cite{LM98} and \cite{MPPT} on the
basis of the explicit expression of the partition function.

A necessary and sufficient condition for the internal energy
$U_{q}(T,V)$ to be proportional to the volume $V$ and to obey
Eq. (\ref{5a}) is that $Z_{q}(A_{d})$ must have the general form
\begin{equation}\label{asch}
Z_{q}(A_{d})=e_q^{C_1(q)A_d}
\end{equation}
where $e_q^x =[1+(1-q)x]^{1/(1-q)}$ is the inverse function of
$\ln_q(x)$ and $C_1(q)$
is an unknown, regular at $q=1$, function. Indeed the relation
(\ref{asch}) cannot be exact.
It can be obtained only as an approximation and $C_1(q)$ depend upon the
used approximation scheme. For example, within the framework of the
factorization approximation used in Ref. \cite{BSD02}, we have
$C_1(q)=[(4-3q)(3-2q)(2-q)]^{-1}$.
Now, though $q\neq1$ the Stefan-Boltzmann law temperature behavior in
its usual form is preserved. The constant $\sigma=\sigma(q)$ must be
$q$-dependent (see e.g. Refs. \cite{TBL95,PPV,WNM,BSD02}).

The same relation as Eq. (\ref{slt}) between $p_{q}(T,V)$ and
$U_{q}(T,V)$ exists in the general case of the $q$-thermodynamics
\cite{com2}.  Here however instead of Eq.~(\ref{eos}) the more general
relation (\ref{imp}) takes place and it is necessary to consider the
following partial differential equation for $U_{q}(T,V)$
\begin{equation}\label{pde}
V\left(\frac{\partial U_{q}}{\partial
V}\right)_{T}=\frac{T}{d}\left(\frac{\partial U_{q}}{\partial
T}\right)_{V}-\frac{1}{d}U_{q}.
\end{equation}
This equation has a solution of the type
\begin{equation}\label{rpdf}
U_{q}(T,V)=\sigma_{q}(T^{d}V)V^{C(q,d)/d}T^{1+C(q,d)},
\end{equation}
where the constant $C(q,d)$ (independent of $T$ and $V$) and the
unknown function $\sigma_{q}(x)$ can be obtained only at the
microscopic level.  This is the generalization of the
Stefan-Boltzmann law that can be obtained without using the
explicit expression for the partition function.

The result (\ref{rpdf}) means that in the considered case we loose the
`famous' $T^{d+1}$ behavior of the internal energy as a function
of the temperature. This is a strict consequence of the fact that
$U_{q}(T,V)$ does not dependent linearly on the volume $V$. This
is in agreement with the findings of Ref. \cite{N03}.
However a consideration on a pure thermodynamic level does not exclude a
$q$-dependence of the proportionality coefficient of the $T^{4}$ law.

Without loss of generality let us consider a system
in a cube with $V=L^{d}$. If we introduce the mean thermal wavelength of
the black-body photons $l=l(T)\equiv\hbar c/kT$, Eq. (\ref{rpdf})
may be transformed into the following scaling forms
\begin{equation}\label{lasteq}
U_{q}(T,L)=\frac{\kappa}{l}\textrm{g}_{q}\left(\frac{L}{l}\right),
\end{equation}
where $\kappa$ is a dimensionless constant and $\textrm{g}_{q}(x)$
is a function, of which the explicit form depends on the way of writing 
the energy constraint (see Ref. \cite{TMP98}) {\it i.e.} its expression 
may be quite different as a function of the ratio $L/l$ 
(see e.g. Eqs.(\ref{asbl}) and (\ref{s2a})) depending on the used 
approach: unnormalized or normalized.

\section{The Wright function}\label{wrightsec}
In Section \ref{introduction} we have
advanced that for the investigation of the black - body radiation
in the context of NSM different approaches has been used in the
literature. Earlier, using the normalized approach the exact $q$
counterpart of (\ref{1}) is found to be \cite{LM98}
\begin{equation}\label{3}
Z_{q}(A_{d})=\Gamma\left(\frac{2-q}{1-q}\right)
\sum_{m=0}^{\infty}\frac{A_{d}^{m}}{(1-q)^{dm}m!}
\frac{1}{\Gamma[(2-q)/(1-q)+dm]}.
\end{equation}
Later, another expression for the partition
function was obtained within the framework of the OLM and the
normalized approaches. It is given \cite{MPPT} by the relation
\begin{equation}\label{3a}
\bar{Z}_{q}(U_{q},A_d)=\Gamma\left(\frac{2-q}{1-q}\right)
\sum_{m=0}^{\infty}\frac{A_{d}^{m}}{(1-q)^{dm}m!}
\frac{[1+(1-q)kT U_{q}]^{dm+1/(1-q)}}{\Gamma[(2-q)/(1-q)+dm]}.
\end{equation}
obtained under the cut-off-like condition
$1+(1-q)(kT)^{-1} U_{q}>0$,
otherwise we have $\bar{Z}_{q}(U_{q},A_d)=0$.

In spite of the fact that in the last case all the thermodynamic
quantities, e.g. the internal energy, can be expressed in an exact
fashion a complication arises. The corresponding expressions are {\it
self-referential} \cite{MPPTa,MPPT} in the sense that the
thermodynamic functions are not expressed in a closed form. This fact
leads to mathematical difficulties that make the problem for the most
part only numerically tractable. Notice that Eqs. (\ref{3}) and
(\ref{3a}) are valid only for $q<1$.

If one tries to apply the above results to the experimental data of
the cosmic back-ground radiation the condition  $A_d\gg1$ is
always satisfied since $\hbar c/(kT)$ is of the order of $1/10$ cm
and $V$ is of cosmological dimensions \cite{N03}.
This physical fact will lead to a great
simplification in the mathematical expressions given in Eqs.
(\ref{3}) and (\ref{3a}). Having this in mind, we take advantage
of the fact that
the series in the r.h.s of Eqs. (\ref{3}) and (\ref{3a}) can be
presented in terms of the entire function
\begin{equation}
\phi (\rho,\alpha;z)=\sum_{m=0}^{\infty}\frac{z^{m}}{m!\Gamma(\rho
m +\alpha)},\qquad \rho >0,\alpha \in \mathbb{C},
\end{equation}
introduced in 1933 by E.M. Wright in the asymptotic theory of
partitions. For analytical properties, some generalizations and
applications of this function the interested reader may consult
Ref. \cite{GLM99}. We note here an useful mathematical result
concerning the behavior of Wright function $\phi(\rho,\alpha;z)$.
If $\rho>0$, for a large real $z$, we have the asymptotic
expansion \cite{GLM99}:
\begin{eqnarray}\label{4}
\phi (\rho,\alpha;z)&=&(\rho z)^{\frac{(1-2\alpha)}{(2+2\rho)}}
\sqrt{\frac{2\pi}{\rho+1}}
\exp[(1+\rho^{-1})(\rho z)^{\frac{1}{(1+\rho)}}]
\nonumber\\[.5cm]
&&\times\left[1+\sum_{m=1}^{M}\frac{(-1)^{m}
a_{m}(\rho,\alpha)}{(\rho z)^{\frac{m}{(1+\rho)}}}+O((\rho
z)^{-\frac{M+1}{(1+\rho)}})\right],
\end{eqnarray}
i.e. the asymptotic behavior of the Wright function is presented
in terms of elementary functions. This result permits to obtain
the different thermodynamic functions of the black-body radiation
in a more simple form, namely, in some particular cases. The constants
$a_{m}(\rho,\alpha)$ can be exactly evaluated \cite{GLM99}. For
our analysis below we need
$$
a_1(\rho,\alpha)=\frac1{\rho+1}\left[\frac\alpha2(\alpha-\rho-1)
+\frac1{24}(2+\rho)(1+2\rho)\right]
$$
and
\begin{eqnarray*}
a_2(\rho,\alpha)&=&\frac1{(1+\rho)^2}\left[\frac\alpha{48}(\alpha-\rho-1)
[6\alpha^2+\alpha(2-14\rho)+\rho(6\rho-7)-2]\right.\\
&& \ \ \ \ \left.+\frac7{1152}(2+\rho)^2[103+4\rho(7+\rho)]\right].
\end{eqnarray*}

\section{The Stefan-Boltzmann law}\label{secsb}
In order to define the unknown function and constants in the
thermodynamic relations discussed in Section \ref{sec2} and to obtain
the $q$-generalization of the Stefan-Boltzmann law one must use the
concrete expression for the partition functions: (\ref{3}) for
$Z_{q}(A_{d})$, obtained using the unnormalized approach, or (\ref{3a})
for $\bar{Z}_{q}(U_{q},A_d)$, which is a result of the normalized
approach. This motivates us to consider below both approaches
separately.

\subsection{Unnormalized approach}
Within this approach the thermodynamic quantities are computed using the
so-called unnormalized expectation values introduced in Eq. (\ref{CT}).
In this case the generalization of the Stefan-Boltzmann law in terms of
the Wright function is given by the following expression for the
internal energy
\begin{equation}\label{5}
U_{q}(T,V)=\frac{dkTA_{d}}{(1-q)^{d}}
\left[\Gamma\left(\frac{2-q}{1-q}\right)\right]^{1-q}\frac{\phi
\left(d,\frac{2-q}{1-q}+d;\frac{A_{d}}{(1-q)^{d}}\right)}{\left[\phi
\left(d,\frac{2-q}{1-q};\frac{A_{d}}{(1-q)^{d}}\right)\right]^{q}}.
\end{equation}

Now, let us consider the physically interesting case $d=3$. In the
limit $q\to1$ the asymptotic expansion (\ref{4}) fails. Then if
$(1-q)$ is fixed, for $A_{3}\gg1$, using Eq. (\ref{4}) (up to the
zeroth order in small values of $z^{-1}$) we get
\begin{equation}\label{asbl}
U_{q}(T,V)=\frac{kT}{(8\pi)^{(1-q)/2}}
\left[\Gamma\left(\frac{2-q}{1-q}\right)\right]^{1-q}
\left[\frac{3A_{3}}{(1-q)^{3}}\right]^{-(1-q)/8}
\exp\left\{\frac{4}{3} \left[3A_{3}(1-q)\right]^{1/4}\right\}.
\end{equation}
The last equation is in full consistency with the relation
(\ref{rpdf}) if for the constant $C(q,3)$ we take the value
$-\frac{3}{8}(1-q)$ and the function $\sigma_{q}(T^{3}V)$ is equal
to the exponential function in the r.h.s. of Eq. (\ref{asbl}) with
the corresponding factor.

\begin{figure}
\resizebox{3.5in}{!}{\includegraphics{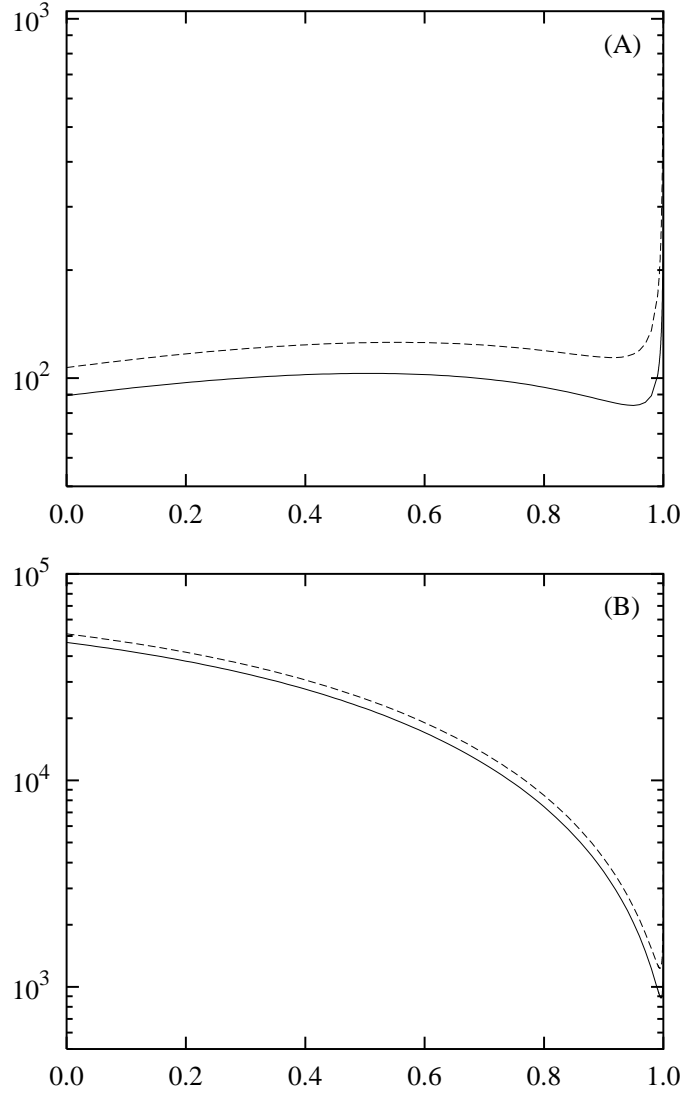}}
\caption{Behavior of $q$-dependence of $U_q(T,V)/3kT$ from Eqs.
(\ref{5}) (solid line) and (\ref{asbl}) (dashed line) for (A) $A_3=500$,
(B) $A_3=5000$. Notice that we used the
logarithmic scale along the vertical axis.}\label{fig1}
\end{figure}

In FIG. \ref{fig1} we present the comparison of the behaviors of
$U_q(T,V)$ from Eqs. (\ref{5}) and (\ref{asbl}) at large $A_3$. It
shows that for large $A_3$, the expression (\ref{5}) for the
internal energy is a good approximation of the exact one given by
(\ref{asbl}). To our knowledge such kind of approximations is
presented for the first time for the black-body radiation problem,
while expansions around $q=1$ investigating deviations from the
BGS are known \cite{TT00}.

As a conclusion we find that our result (\ref{asbl}) shows that in the
case of unnormalized approach the Stefan-Boltzmann law is not
preserved. Remark that in this case the internal energy is not
proportional to the volume. This is in concert with the discussion of
Section \ref{sec2}.

\subsection{Normalized approach }
Within this framework the expectation values are computed using Eq.
(\ref{OLM}). This approach has the advantage of reproducing the
traditional thermodynamic relations.
The internal energy $U_q\equiv U_q(T,V)$ is found to obey a nonlinear
equation that can be expressed in terms of the Wright function.
It has the form:
\begin{equation}\label{eq}
U_{q}=kT\frac{dA_{d}[1+(1-q)(kT)^{-1}
U_{q}]^{d+1}}{(1-q)^{d+1}}\frac{\phi
\left(d,\frac{2-q}{1-q}+d;\frac{A_{d}[1+(1-q)(kT)^{-1}
U_{q}]^{d}}{(1-q)^{d}}\right)}{\phi
\left(d,\frac{1}{1-q};\frac{A_{d}[1+(1-q)(kT)^{-1}
U_{q}]^{d}}{(1-q)^{d}}\right)}.
\end{equation}
For fixed $(1-q)$ and
\begin{equation}\label{asmp2}
\frac{A_{d}[1+(1-q)(kT)^{-1} U_{q}]^{d}}{(1-q)^{d}}\gg1,
\end{equation}
using Eq. (\ref{4}) (up to the first order in small values of $z^{-1}$)
after some algebra Eq. (\ref{eq}) reads
\begin{eqnarray}
(1-q)(kT)^{-1} U_q &=& [1+(1-q)(kT)^{-1} U_q]\nonumber\\
&&\times\left[1-\frac1{1-q}\left(\frac{dA_d}{(1-q)^d}
\left[1+(1-q)(kT)^{-1} U_q\right]^d\right)^{-\frac1{1+d}}\right].
\end{eqnarray}
The solution of this equation is surprisingly simple. The result is
\begin{equation}\label{s2a}
U_q=\sigma VT^{d+1}-\frac1{1-q}kT,
\end{equation}
where
$$
\sigma=\frac{\Gamma(d)\zeta(d+1)2d(d-1)}{(4\pi)^{d/2}\Gamma(d/2)}
\frac{k^{d+1}}{(\hbar c)^d}
$$
is the {\it usual} Stefan-Boltzmann constant. Since it is not possible
to take the limit $q\to1$ in Eq. (\ref{s2a}) we cannot recover the
Stefan-Boltzamnn any more. Inserting now Eq. (\ref{s2a}) into Eq.
(\ref{asmp2}) we obtain the condition
\begin{equation}\label{ner} A_{d}(dA_{d})^{d}\gg1,
\end{equation}
restricting the range of the validity of the obtained solution for
$U_{q}$. In order to improve our result we calculated the next term
of the internal energy (\ref{s2a}). To this end we used the
expansion of r.h.s of (\ref{eq}) to the second term i.e. in small
values of $z^{-2}$. The complicated ensuing equation could be solved
using an iteration method, which leads to the additional term
$[1-(1-q)d][2d(1-q)]^{-1}kT$ in the solution (\ref{s2a}). Note here that
in spite of taking into account the next order in our calculations we see
that our results are not improving in the sense that we cannot take the
limit $q\to1$.

Remark that the internal energy (\ref{s2a}) is noextensive
for relatively small volumes of the system. In the thermodynamic
limit it becomes extensive, confirming the conclusions of Ref.
\cite{abe1999}. The entropy, $S$, follows from the thermodynamic
relation $(\partial S/\partial V)_T=(\partial p/\partial T)_V$. It
has been demonstrated that in the case of the NSM the traditional
thermodynamic relations remains valid only if one uses the R\'enyi
entropy, $S^R$, instead of its Tsallis counterpart $S_q^T$
\cite{abe2001}. The first one has the remarkable property of being
extensive. In the thermodynamic limit we have in the case under
consideration
\begin{equation}\label{renyi}
S^R=V\sigma\frac{d}{d+1}T^d.
\end{equation}
The Tsallis entropy can be deduced through the relation
\cite{tsallis2004}
\begin{equation}\label{tsallis}
S_q^T=\frac k{1-q}(\exp\left[(1-q)S^R\right]-1).
\end{equation}
This is a sign that the Stefan-Boltzmann law remains valid in the NSM
context as well.

\section{Discussion}\label{discussion}

Let us discuss the most important case $d=3$.
In our consideration a crucial point is the condition
\begin{equation}\label{2a}
A_{3}=\frac{\pi^{2}}{45} \left(\frac{\hbar
c}{kT}\right)^{3}V\gg1
\end{equation}
that was used to truncate the asymptotic expansion (\ref{4}) for
obtaining the results given by Eq. (\ref{asbl}) or Eq. (\ref{s2a}).

Our consideration shows that the application of the thermodynamical
concepts of the NSM may lead to the $T^{4}$ Stefan-Boltzmann law. This
takes place if the partition function has the form (\ref{asch}).  This
form would be a result of some approximations (see e.g. \cite{BSD02})
and the $T^{4}$ behavior is a strict consequence of a linear on $V$
dependence of $U(T,V)$. What is important to note is that the
inequality (\ref{2a}) prohibits the use of any formula of the type of
Eq. (\ref{asch}) to the {\it cosmic back-ground} radiation.

The criticism in Refs. \cite{N03,Ncm} against the use of the
formula obtained under the condition that $A_{3}$ is considered as
a small parameter and after that utilized to fit the
data of the cosmic black-body radiation is supported by our
consideration \cite{com2a}. If one tries to apply the above results to
the experimental data of the cosmic back-ground radiation the
condition (\ref{2a}) is always satisfied as it was mentioned in
Section \ref{wrightsec} since $\hbar c/(kT)$ is of the order of
$1/10$ cm and $V$ is of cosmological dimensions. On the other
hand, as it is mentioned fairly in the reply \cite{T03} the
criticism of Ref. \cite{N03} concerns on equal ground both
approaches: unnormalized and normalized which does not clarify the
problem. In Ref. \cite{T03} it is suggested that the last one is
free of inconsistencies argued in Ref. \cite{N03}. In our
understanding at this point the situation seems to be
clear \cite{com3}. Here we, consider in more details both 
approaches unnormalized and normalized separately. 

In the case of the normalized approach our
investigation results in the formula (\ref{asbl}). If the inequality
(\ref{2a}) is fulfilled we loose the $T^{4}$ behavior and in addition
the {\it energy density}, Eq. (\ref{asbl}), depends on the
volume, which remains unacceptable. This agrees with the results of
Ref.~\cite{N03}.

Free of such a defect would be a theory based on the normalized
approach \cite{MPPT,TMP98,T03}. In this case one can immediately
see that the condition (\ref{ner}) is the relaxed version of
(\ref{2a}). This means that if one tries to apply the normalized
approach to the analysis of the cosmic back-ground radiation the
expression (\ref{s2a}) has to be used. This result is consistent
with the thermodynamic relations (\ref{pde}) and (\ref{rpdf}).
Indeed the first term in Eq. (\ref{s2a}) is the usual
Stefan-Boltzman law. The question is: how to interpret the last
one? The wisdom of the standard statistical mechanics is that such
terms are to be omitted since they are of the order of $O(1/V)$
and do not contribute in the thermodynamic limit. On the other
hand, this term diverges with $q\rightarrow 1^-$ and may be
considered as a sign that in NSM the Boltzman-Gibbs limit
$q\rightarrow 1$ and the thermodynamic limit do not commute with
each other. This is in agreement with the results obtained in the
framework of a classical gas \cite{A99}. In order to avoid any
misunderstanding, let us emphasize that our treatment excludes any
attempt to take the limit $1-q$ and indeed $U_q>0$.

\begin{acknowledgments}
The authors thank J. G. Brankov and D. I. Pushkarov for useful comments and
discussions.
\end{acknowledgments}


\begin{thebibliography}{00}
\bibitem{com2} It is easy to see this from Eqs. (\ref{U}) and
(\ref{P}). Indeed this formalism  is supposed to satisfy the laws of
thermodynamics. Otherwise it would be nonsensical.
\bibitem{com2a} In Refs. \cite{TBL95,PPV,AT01} the  data
have been analyzed in search of possible deviation from
extensivity.
\bibitem{com3} The critics of Refs. \cite{N03,Ncm} concerns a group of
problems. Here, we discuss only those related to applicability
of $T^{4}$ law and its possible generalization to cosmic
back-ground radiation.
\end{thebibliography}
\end{document}